\documentclass[fleqn,3p,onecolumn,12pt]{consistency-modmax_elsarticle}  

\newcommand{\versionnumber}{v4}  

\journal{Nuclear Physics B 856 (2012) 666}  
\renewcommand{\today}{arXiv:1110.4101 (\versionnumber)}

\usepackage{graphicx}\usepackage{bm}
\usepackage{mathrsfs}\usepackage[latin1]{inputenc}
\usepackage{stmaryrd}\usepackage{array}
\usepackage{psfrag}\usepackage{dsfont}
\usepackage{epsfig}\usepackage{titletoc}
\usepackage{float}   
\usepackage{wrapfig} 
\usepackage{amsmath}\usepackage[latin1]{inputenc}
\usepackage{amsmath}\usepackage{amssymb,stmaryrd}
\usepackage{mathrsfs}\usepackage{array}
\usepackage{graphicx}\usepackage{psfrag}
\usepackage{dsfont}\usepackage{epsfig}\usepackage{cancel}

\begin{document}

\begin{frontmatter}

\title{Models for low-energy Lorentz violation in the photon
sector: Addendum to `Consistency of isotropic modified Maxwell theory'}

\author{F.R. Klinkhamer\corref{cor1}}
\cortext[cor1]{Corresponding Author}\ead{frans.klinkhamer@kit.edu}
\author{M. Schreck} \ead{marco.schreck@kit.edu}
\address{Institute for Theoretical Physics, University of Karlsruhe,\\
         Karlsruhe Institute of Technology, 76128 Karlsruhe, Germany}

\begin{abstract}
In a previous article~\cite{KlinkhamerSchreck2011},
we established
the consistency of isotropic modified Maxwell theory
for a finite range of the Lorentz-violating parameter
$\widetilde{\kappa}_{\mathrm{tr}}$, which includes both positive and
negative values of $\widetilde{\kappa}_{\mathrm{tr}}$.
As an aside, we mentioned the existence of
a physical model which, for low-energy photons, gives rise to isotropic
modified Maxwell theory with a positive parameter
$\widetilde{\kappa}_{\mathrm{tr}}$ (corresponding to a ``slow'' photon).
Here, we present a related  model which gives rise to isotropic
modified Maxwell theory with a negative parameter
$\widetilde{\kappa}_{\mathrm{tr}}$ (corresponding to a ``fast'' photon).
Both models have an identical particle content, photon and
Dirac particles, but differ in the type of spacetime
manifold considered.\vspace*{0.5\baselineskip}
\end{abstract}
\end{frontmatter}

\vspace*{1\baselineskip}
We have investigated the consistency of isotropic modified Maxwell theory
in Ref.~\cite{KlinkhamerSchreck2011},
which contains an extensive list of references.
The photon (${\gamma}$) of this theory has the following
dispersion law and velocity:
\begin{subequations}\label{eq:dispersion-law-group-velocity-isotropic-modmax}
\begin{eqnarray}
\label{eq:dispersion-law-isotropic-modmax}
\omega_{\gamma}(\mathbf{k})&\equiv& c\,|k^{0}(\mathbf{k})| =
\sqrt{\frac{1-\widetilde{\kappa}_{\mathrm{tr}}}
           {1+\widetilde{\kappa}_{\mathrm{tr}}}}\;\,c\,|\mathbf{k}|=
\Big(1-\widetilde{\kappa}_{\mathrm{tr}}
+\mathsf{O}\big(\widetilde{\kappa}_{\mathrm{tr}}^{\,2}\big)\Big)\;c\,|\mathbf{k}|\,,
\\[2mm]\label{eq:group-velocity-isotropic-modmax}
v_{\gamma,\;\text{group}}(\mathbf{k})&\equiv&
\left|\frac{\partial\,\omega_{\gamma}(\mathbf{k})}{\partial\,\mathbf{k}}\right|
=
\sqrt{\frac{1-\widetilde{\kappa}_{\mathrm{tr}}}
           {1+\widetilde{\kappa}_{\mathrm{tr}}}}\;c=
\Big(1-\widetilde{\kappa}_{\mathrm{tr}}
+\mathsf{O}\big(\widetilde{\kappa}_{\mathrm{tr}}^{\,2}\big)\Big)\;c\,.
\end{eqnarray}
\end{subequations}
Here, $\hbar\,\mathbf{k}$ is the three-momentum of the photon,
$\hbar\,\omega_{\gamma}$ the energy of the photon,
$\widetilde{\kappa}_{\mathrm{tr}}$ a dimensionless model parameter
(coupling constant) characterizing the strength of the Lorentz
violation, and
$c$ the velocity of light in vacuum without Lorentz violation
(a better definition will be given shortly).
The main result of Ref.~\cite{KlinkhamerSchreck2011}
was that isotropic modified Maxwell theory is consistent
(i.e., causal and unitary)
for model  parameters $\widetilde{\kappa}_{\mathrm{tr}}\in (-1,1]$.

The question arises if there exists an underlying physical model which,
at low photon energies, gives rise to an effective theory containing
isotropic modified Maxwell theory. In Ref.~\cite{KlinkhamerSchreck2011},
we have briefly discussed a model~\cite{BernadotteKlinkhamer2007}
which gives isotropic
modified Maxwell theory with $\widetilde{\kappa}_{\mathrm{tr}}>0$
(corresponding to a ``slow'' photon with velocity below $c$).
Here, we present another model which gives isotropic
modified Maxwell theory with $\widetilde{\kappa}_{\mathrm{tr}}<0$
(corresponding to a ``fast'' photon with velocity above $c$).

Both models have the same particle content but different
spacetime manifolds.
The particle theory is, in fact, a trivial extension of standard
quantum electrodynamics (QED) with photons,
charged Dirac particles (``electrons''),
and uncharged Dirac particles (``neutrinos'')
propagating in a background spacetime.
There are thus three quantum fields with the following physical
charges and masses of the corresponding particles:
\begin{subequations}\label{eq:QED-fields}
\begin{eqnarray}
A_{\mu}(x)    &:& Q=0\,,   \qquad\;\, M=0 \,, \\[2mm]
\psi_{e}(x)   &:& Q=-|e|\,,\quad M=m_{e} \,,\\[2mm]
\psi_{\nu}(x) &:& Q=0\,,   \qquad\;\, M=m_\nu \,,
\end{eqnarray}
\end{subequations}
to which is added the classical background field $g_{\mu\nu}(x)$.
This relatively simple theory allows for a clean operational
definition of $c$,
namely, as the maximal attainable velocity of the neutrino
in a local orthonormal frame:
\begin{subequations}\label{eq:def-c-neutrino-disp-law}
\begin{eqnarray}\label{eq:def-c}
c &\equiv&
\left[\;\lim_{|\mathbf{k}|\,\to\,\infty}\;
\frac{\omega_\nu(\mathbf{k})}{|\mathbf{k}|}\;
\right]^{\text{(local)}}\,,
\\[2mm]\label{eq:neutrino-disp-law}
\big(\omega_\nu(\mathbf{k})\big)^2 &=&
c^2\,|\mathbf{k}|^2+ \big(m_\nu\,c^2/\hbar\big)^2 \,.
\end{eqnarray}
\end{subequations}
Even without weak interactions, it is still possible, in principle,
to monitor the propagation of a neutrino wave packet by its
gravitational effects on test masses.

The first model, now, considers these quantum fields to propagate over
a certain type of classical spacetime-foam manifold,
which is flat over large
scales but has small localized defects (``holes'').
The propagation of standard
electromagnetic waves in different realizations of such a spacetime foam
has been investigated in Ref.~\cite{BernadotteKlinkhamer2007}.
The result is a modified photon dispersion law
in the long-wavelength limit, which can be compared
to the dispersion law \eqref{eq:dispersion-law-isotropic-modmax}
of isotropic modified Maxwell theory.
The parametric dependence of the effective Lorentz-violating
coupling constant $\widetilde{\kappa}_{\mathrm{tr}}$
turns out to be as follows:
\begin{equation}\label{eq:kappatildetrace-parametric-foam}
\widetilde{\kappa}_{\mathrm{tr}}\sim \big(\,b\big/l\,\big)^4\,,
\end{equation}
with $b$ the typical size of a spacetime defect
and $l$ the mean separation.
The calculation~\cite{BernadotteKlinkhamer2007} assumes $l \gg b$
and uses the long-wavelength limit $\lambda_\text{photon}\gg l$.

Physically, a positive Lorentz-violating parameter
$\widetilde{\kappa}_{\mathrm{tr}}$ can thus be interpreted
as the spacetime-volume fraction excluded by
defects.\footnote{A spacetime manifold with punctures
($b \to 0$ for fixed $l$)
may still give a finite positive value for the effective coupling
constant $\widetilde{\kappa}_{\mathrm{tr}}$.
The particle content then needs to be changed to that of
a chiral gauge theory, having, for example, a $U(1)$ gauge boson
with massless Weyl fermions in a ``complex'' representation
(appropriate charges proportional to $e$).
Anomalous effects then give a positive parameter
$\widetilde{\kappa}_{\mathrm{tr}} \propto(e^2/4\pi)^2 \equiv \alpha^2$,
according to Eqs.~(5.26)--(5.28) of Ref.~\cite{KlinkhamerRupp2004}.
Incidentally, this quadratic dependence on $\alpha$
corrects the erroneous statement in
Footnote.~4
of Ref.~\cite{KlinkhamerSchreck2011} about a linear dependence.}
An equally important result from
Ref.~\cite{BernadotteKlinkhamer2007} is that the dispersion law of
Dirac fermions is not modified to leading order.
Hence, the model neutrino propagating over this type
of classical spacetime foam keeps the standard dispersion law
\eqref{eq:neutrino-disp-law}, and similarly for the model electron.
The effective theory from this first model thus corresponds to the
isotropic modified Maxwell theory with $\widetilde{\kappa}_{\mathrm{tr}}> 0$
and the standard Dirac theory for the fermions.
Concretely, the model has an effective photon velocity which is
smaller than the maximum velocity $c$ of the neutrino and electron.

The second model considers the same quantum fields to propagate over
a smooth simply-connected spacetime manifold with large-scale curvature.
Drummond and Hathrell~\cite{DrummondHathrell1980}
(see also Ref.~\cite{Shore2001} for further discussion)
have already examined the
propagation of photons in a gravitational background,
taking into account the effects
of vacuum polarization by the electron/positron field.
Again, the calculation uses a long-wavelength limit (see below).
Notably for a Friedmann--Robertson--Walker
(FRW) universe with nonvanishing matter content
(energy density $\rho$ and pressure $P$),
they found that photons propagate
with a speed that is both isotropic and larger than
the speed of light in flat Minkowski spacetime.
Translated to a model with particle content \eqref{eq:QED-fields},
the result is that photons propagate in FRW spacetime
with a speed that is both isotropic and larger than
the maximum attainable speed of the uncharged Dirac particle (neutrino).
Conceptually, it is, of course, better to compare velocities
in the same spacetime.

The second model with an FRW background metric thus gives
rise to an effective theory~\cite{DrummondHathrell1980}
with the linearized photon dispersion law
\eqref{eq:dispersion-law-isotropic-modmax}
of isotropic modified Maxwell theory
for negative $\widetilde{\kappa}_{\mathrm{tr}}$.
Reinstating $\hbar$ and $c$ in Eqs.~(2.19) and (6.8)
of Ref.~\cite{DrummondHathrell1980} gives
\begin{equation}\label{eq:k0-veckabs-ratio}
\frac{|k^{0}|}{|\mathbf{k}|}=
1+\frac{11}{45}\,\alpha\,\left(\frac{\hbar}{c^3}\right)^2
\frac{1}{(m_{e})^2}\;G_{N}\,\big[\,\rho(t)+P(t)\,\big]\,,
\end{equation}
with fine-structure constant
$\alpha\equiv e^2/(4\pi\hbar c)$ and cosmic time $t$ from the FRW metric.
The velocity $c$ in
\eqref{eq:k0-veckabs-ratio} corresponds to the maximum neutrino
velocity \eqref{eq:def-c}, as will be discussed further below.

In order to get an estimate for the magnitude of this result,
consider the cold-dark-matter component of the present
(spatially-flat) universe with pressure $P=P_\text{CDM}=0$
and an energy density $\rho=\rho_\text{CDM}$ equal to approximately
one quarter of the critical energy density
$\rho_c\equiv 3\,c^2\,H_0^2/(8\pi\,G_{N})$
for Hubble constant $H_0$.
With these approximations, \eqref{eq:k0-veckabs-ratio} becomes
\begin{equation}\label{eq:k0-veckabs-ratio-estimate}
\left[\,\frac{|k^{0}|}{|\mathbf{k}|}\,\right]^{\text{(present\;universe)}}
\simeq
1+\frac{11}{180}\,\alpha\,\left(\frac{\hbar}{m_{e}\,c}\right)^2\,
\frac{G_{N}\,\rho_c}{c^4}
\simeq 1+\frac{11}{480\,\pi}\,\alpha\,
\left(\frac{\hbar}{m_{e}\,c}\right)^2\, \frac{H_0^2}{c^2}\,.
\end{equation}
The parametric dependence of the corresponding Lorentz-violating
parameter $\widetilde{\kappa}_{\mathrm{tr}}$
can then be written as follows~\cite{DrummondHathrell1980}:
\begin{subequations}\label{eq:kappatildetrace-parametric-def-lambdaC-LHubble}
\begin{eqnarray}\label{eq:kappatildetrace-parametric-curvature}
\widetilde{\kappa}_{\mathrm{tr}}&\sim& -
\alpha \,\Big(\lambda_{e,\,\mathrm{Compton}}/L_{\mathrm{Hubble}}\Big)^2\,,
\\[2mm]
\lambda_{e,\,\mathrm{Compton}}&\equiv& \hbar/(m_{e}\,c),\,
\quad
L_{\mathrm{Hubble}}\equiv c/H_0\,.
\end{eqnarray}
\end{subequations}
Hence, we have expressed the effective Lorentz-violating
parameter by the fundamental length scales that play a role in this setup:
the reduced Compton wavelength of the electron and the
Hubble length, which is the curvature scale of the cosmological
background spacetime.\footnote{In the present universe
with $H_0 \sim 70\;\text{km}\,\text{s}^{-1}\,\text{Mpc}^{-1}$,
$L_{\mathrm{Hubble}} \sim 10^{26}\;\text{m}$, $\alpha\sim 10^{-2}$,
and $m_{e} \sim 0.5\;\text{MeV}$,
the numerical value of \eqref{eq:kappatildetrace-parametric-curvature}
is of order $-10^{-79}$,
which may be compared to
the current experimental bound at the $-10^{-15}$
level~\cite{KlinkhamerSchreck2008}.}
As mentioned above, the calculation of Ref.~\cite{DrummondHathrell1980}
uses the long-wavelength limit, specifically
$\lambda_\text{photon}\gg \lambda_{e,\,\mathrm{Compton}}$,
but the calculation
still keeps $\lambda_\text{photon}\ll L_{\mathrm{Hubble}}\,$.

The fine structure constant appears in
\eqref{eq:k0-veckabs-ratio}--\eqref{eq:kappatildetrace-parametric-def-lambdaC-LHubble}
since the effective Lorentz violation results from a
one-loop quantum effect. The electron contributes to the quantum fluctuation
of the photon and gives it a size of the order of the electron Compton wavelength.
Since the photon acquires a size by quantum fluctuations, the equivalence
principle is not exactly valid for this particle,
which is no longer pointlike. The same holds for the
electron but not for the neutrino (at least, the neutrino of the model
considered, with a photon but not a $Z$ boson).

The effective theory from this second model thus corresponds to
isotropic modified Maxwell theory
with $\widetilde{\kappa}_{\mathrm{tr}}< 0$
and standard Dirac theory for the neutrino.
Concretely, the model has an effective photon velocity which is
larger than the maximum velocity $c$ of the neutrino.\footnote{To
leading order in $\alpha$, there is, however,
no Lorentz violation for a photon propagating in a de Sitter
background~\cite{DrummondHathrell1980,Shore2001}, giving
$\widetilde{\kappa}_{\mathrm{tr}}=0$ in the effective
modified Maxwell theory.
This result is consistent with \eqref{eq:k0-veckabs-ratio},
because the corresponding Lorentz-invariant vacuum energy
and pressure are related by $\rho_V+P_V=0$.}

It is clear that both models can be combined:
the QED fields \eqref{eq:QED-fields} now propagate over a
spacetime manifold with both small-scale structure and
large-scale curvature. Depending on the relative strength of
\eqref{eq:kappatildetrace-parametric-foam}
and \eqref{eq:kappatildetrace-parametric-curvature},
the effective photon velocity will then be
below or above the maximum neutrino velocity $c$.

To summarize, there exist underlying physical models
which, in the long-wavelength limit of the photon,
give rise to isotropic modified Maxwell theory
with both $\widetilde{\kappa}_{\mathrm{tr}}\geq 0$
and $\widetilde{\kappa}_{\mathrm{tr}}\leq 0$.
But these models generate the isotropic-modified-Maxwell
dispersion law only to linear order
in $\widetilde{\kappa}_{\mathrm{tr}}$.
In other words, isotropic modified Maxwell theory is obtained
for sufficiently small values of $|\widetilde{\kappa}_{\mathrm{tr}}|$.
This may explain the consistency of isotropic modified Maxwell theory
for small values of the Lorentz-violating parameter.
The consistency for \emph{finite} parameter values
$\widetilde{\kappa}_{\mathrm{tr}}\in (-1,\,1]$ is, therefore,
a nontrivial result~\cite{KlinkhamerSchreck2011}.



\end{document}